\documentstyle[aasms4,12pt,epsf]{article}

\begin{document}

\title{ Near Infrared Observations of the  Extremely Red Object
CL~0939+4713~B : An Old 
Galaxy at z$\sim$ 1.58 ? \footnote{Based on observations obtained 
at the W.M. Keck Observatory, 
which is operated as a scientific partnership among the California 
Institute of Technology, the
 University of California and the National Aeronautics and Space 
Administration}}

\author{B.T. Soifer\altaffilmark{2,3},  K. Matthews\altaffilmark{2}, 
G. Neugebauer\altaffilmark{2}, L. Armus\altaffilmark{3},J. G.
Cohen\altaffilmark{4}, 
S.E. Persson\altaffilmark{5}, I. Smail\altaffilmark{6}}

\altaffiltext{2}{Palomar Observatory, 320-47, Caltech, Pasadena, CA 
91125}

\altaffiltext{3}{SIRTF Science Center, 310-6, Caltech, Pasadena, CA 
91125}

\altaffiltext{4}{Palomar Observatory, 105-24, Caltech, Pasadena, CA 
91125}
\altaffiltext{5}{Carnegie Observatories, 813 Santa Barbara St. Pasadena, 
CA 91101}
\altaffiltext{6}{Department of Physics, University of Durham, South Road,
Durham DH1 3LE, UK}

\received{1999 August}

\begin{abstract}

Near infrared imaging and spectroscopic observations of the  extremely
red object ($R-K \sim 7$ mag) CL~0939+4713~B  have been obtained with
the Near Infrared Camera  on the Keck I Telescope of the W. M. Keck
Observatory. The imaging shows a slightly elongated structure, while
the  spectroscopy shows a continuum break   that allows us to determine
the redshift of z $= 1.58 + 0.01/-0.03$ for this system.  The fits of a
range of models to the infrared spectrum suggests that it is
predominantly  an old ($> 10^9$ yrs) stellar system that suffers
little  extinction, while the measurerd R and I magnitudes suggests an age of
$\sim 3 \times 10^8 $ years.  The limit on the equivalent width of any
emission line in the infrared spectrum argues that CL~0939+4713~B is
not an actively star forming galaxy.  This system, though similar in
$R-K$ color to HR 10 [also known as J1645+46] (Dey et al. 1999), is
much different in morphology and emission line strengths, demonstrating
the heterogeneity of extremely red extragalactic objects  (EROs)
selected on the basis of large values of $R-K$.

\medskip

\end{abstract}


\section{Introduction}
\label{sec:introduction}

Over the last decade several groups (Elston, Rieke \& Rieke,
1988,1989, Eisenhardt \& Dickinson, 1992, McCarthy, Persson \& West,
1992, Persson, McCarthy, Dressler \& Matthews, 1993, Hu \& Ridgway,
1994, Dey, Spinrad \& Dickinson, 1995, Djorgovski et al. 1995, Barger et al.
1999) have
reported finding a class of extremely red objects  (EROs) with $R-K>$
6 mag that are reasonably bright (K$<$19 mag) and are relatively frequent (surface
densities $\sim 0.01$/sq arcmin at K$<$18 mag, Hu \& Ridgway,
Thompson et al. 1999).

The earliest studies  suggested that these systems  might be at
extremely high redshifts (Elston, Rieke \& Rieke 1988), but this has
not been borne out.  Recent work has shown that these systems have
spectral energy distributions in the visible and near infrared
consistent with elliptical galaxies or very dusty galaxies at z$>$1 
(Persson et al 1993, Hu
\& Ridgway, 1994, Trentham, Kormendy and Sanders, 1999).  
Recently Graham \& Dey (1996) have reported
imaging and spectroscopy of one of the brightest of these red objects,
HR 10,  discovered by Hu \& Ridgway.  Graham and Dey  report the
detection of a spectral feature at 1.6\,$\mu$m which they associate
with H$\alpha$ at a redshift z$=$1.44.  They further suggest that the
morphology of this system is  consistent with an interacting galaxy system, suggesting that this is
a dusty starburst galaxy or AGN.  Dey et al. (1999) confirm the
redshift of HR 10 with  the detection of [OII]3727\AA. Both Dey et
al. and Cimmatti et al. (1998)  report the detection of HR 10 at
submillimeter wavelengths, supporting the suggestion of its being a
dusty, interacting galaxy. Smail et al. (1999) have also suggested that 
up to half of the EROs could be associated with submillimeter
sources detected in SCUBA surveys.

Understanding the nature of EROs, or even whether they represent a
homogeneous class, is important for several reasons.  If these are
indeed  starburst/AGN systems at z$>$1, they could be the high
redshift analogs of the ultraluminous infrared galaxies seen locally
(e.g. Soifer et al. 1984, Sanders et al 1988, Trenthan, Kormendy and 
Sanders, 1999 ).  If powered by hot,
young stars, such systems would have star formation rates greater than
that found for most of the young galaxies at  z$\ge$3  found by
Steidel et al.  (1996).  Alternatively if they are distant passively
evolving elliptical galaxies, then the determination of their ages
might well place  significant constraints on the formation epoch of
massive galaxies.

As part of a program to study the nature of these objects we have
undertaken near infrared spectroscopy of one of the brightest of these
objects  on the Keck I Telescope of the W.M. Keck Observatory.  In
this paper we report spectroscopy and imaging for  CL~0939+4713~B, an
object   serendipitously found  by Persson et al. (1993) in the
infrared in the field of   a rich cluster of galaxies at z=0.41
(Dressler and Gunn, 1992) and referred to by them as $9\alpha  \beta$
B.  It is  non-stellar, and located $\simeq 4''$ southwest of galaxy
174 in the cluster (Dressler and Gunn). Persson et al. reported a $K$
magnitude of 18.3 mag and  $r-K > 7.4$ mag.  Assuming a passively
evolving stellar population, Persson et al.  estimated a  redshift of
1.8 for CL~0939+4713~B  and an age of $3 \times 10^9$ yr, by fitting
their photometry at $r, i, J, H$ and $K$ to Bruzual (1985) models.
The grism observations reported here allow us to search for emission
lines such as those found in the spectrum of HR 10, as well as to
directly measure any continuum in the spectrum. Throughout this paper
we adopt $H_o =$50 km s$^{-1}$Mpc$^{-1}$ and $q_o=0.5$ (and $\Lambda =
0$).

\smallskip

\section{Observations and Data Reduction}
\label{sec:observations}

Photometric and grism observations of CL~0939+4713~B were made on 1997
March 31 using the Near Infrared Camera (NIRC; Matthews and Soifer,
1994) on the Keck I Telescope.  The array has 256 $\times$ 256 pixels
and a scale of 0.15 $''$/pixel.  Photometry  at $J$(1.27\,$\mu$m),
$H$(1.65\,$\mu$m) and $K$(2.2\,$\mu$m) was obtained in  seeing with
FWHM $\sim 0.4''$; the data were calibrated by reference to
calibration stars of Persson et al.(1998).

The final $K$-band image shown in Figure 1 was created by combining
15 individual frames of 12 seconds each, taken both before and after
the spectral data.  The telescope was moved 5-10$''$ between
individual frames, and sky and flat field images were constructed from
the data themselves, after masking out bright sources in the field.
Although there are  no obvious point sources on the image of
CL~0939+4713~B itself, the seeing at the time of the observations was
estimated to have a FWHM $\simeq$ 0.4$''$, measured from standard
stars observed immediately before the data were taken.

Low resolution spectra were obtained that span the range $1 -
2.4$\,$\mu$m in two wavelength settings. For the wavelength range
$1.0-1.6$\,$\mu$m 4500 seconds of observations were made, while 7800
seconds of  observations were made for the range $1.6-2.4$\,$\mu$m;
individual  observations were $\sim$ 300 sec in duration. The slit was
4.5 pixels ($\sim  0.68''$) wide and aligned essentially north-south
so the bright galaxy to the south was included in most spectra.  The
object was moved to five positions  along the slit for successive
integrations.

The data were processed in a conventional manner. The spectra were
corrected  for atmospheric features by  dividing the spectra of Cl
0939+4713~B by that of a G6V star observed on the same night at
similar air mass. The spectrum of the G6V star closely approximates
that of a blackbody of  temperature 5600$K$ at the spectral resolution
of the observations and so was assumed to follow the  blackbody
spectrum.  The infrared spectrum and R band photometry were put on a
common absolute scale by integrating the infrared spectra over the
appropriate wavelength bands and scaling the spectral flux densities
to correspond to the photometry in a 3 $''$ diameter beam. The
infrared data were boxcar smoothed to 2 pixels,
i.e. $\Delta\lambda=$0.0135\,$\mu$m,  the spectral  resolution
corresponding to half the slit width.

In addition to the near infrared observations, images of CL~0939+4713~B
were taken using the Low Resolution Imaging Spectrograph (LRIS; Oke et
al. 1995) on the Keck~II Telescope.  Three 600 second images,  slightly
offset in pointing  location between images,  were taken through an $R$
filter in 1998 March.   The seeing on the summed image was FWHM $\simeq
0.9''$. Eighteen 200 second images with slight positional offsets
were taken in the  $I$ filter  on 1998 November
01 (see Smail et al. 1999). The night was photometric and the total
integration time was 3,600 sec; the seeing was 0.65 $''$ FWHM.  The
data for both images were reduced using standard techniques.  The
photometry was calibrated to the standard system of Landolt (1992).

\smallskip

\section{Results}
\label{sec:results}

The $R$ and $K$ images of CL~0939+4713~B are shown in Figure 1 while
Table 1 reports the photometry for the source.  The $K$ image shows
that CL~0939+4713 B is resolved, having a bright core with a clear
elongation in the east west direction. The $K$ magnitude of CL
0939+4713~B from the Keck imaging is 18.16$\pm$0.06 mag in a 3$''$
diameter beam. To determine whether this value is affected by the
presence of the nearby brighter galaxy (galaxy 174), the bright galaxy
was subtracted from the image by rotating the image by  $180^{\circ}$
about the center of galaxy 174 and subtracting it from the original
image. This successfully eliminated galaxy 174, while preserving the
other objects in the image. The $K$ magnitude of CL~0939+4713~B
measured in the image where galaxy 174 was eliminated is 18.26$\pm$0.10
mag.  Our conclusion from this is that the two measurements are
consistent within the uncertainties and  galaxy 174  does not
significantly affect the measurement of CL~0939+4713~B at $K$ in a
3$''$ diameter beam.  The $K$ measurement of Persson et al. (1993) is
consistent with the value reported here.

The spectrum of CL~0939+4713~B, plotted vs. observed wavelength from
$1.0 - 2.4$\,$\mu$m, is shown in Figure 2.  The flux scale of Figures 2
and 3 is set so that the flux level in the spectrum agrees with the
average of the Keck photometry at  $J, H$ and $K$ as presented in Table
1.   Because the spectral standard that was used is not calibrated
photometrically, there is no independent calibration of the
spectrophotometry.  The spectrum is comparatively  flat  for
wavelengths $\lambda >$ 1.2\,$\mu$m and drops rapidly below  this
wavelength.  There are no strong narrow emission features in the
spectrum, and the only strong continuum feature is the apparent change
of slope at 1.2\,$\mu$m.

In the $R$ image the object is also resolved, with a measured FWHM of
1.27$''$.   The original $r$ magnitude for CL~0939+4713~B reported by
Persson et al. (1993) was $r>25.7$ mag, based on a non-detection of the
source in the $r$ image of the field.  The Keck $R$ image  detects the
source with a measured magnitude of 25.57 mag in a 1.3$''$ diameter
beam.  The aperture correction to a  3$''$ diameter was determined to
be 0.44 mag (with substantial scatter  $\sim$ 0.4 mag) using objects of
similar brightness in the same image.  This leads to a total R
magnitude of 25.13 +0.5/$-$0.3 mag.  Uncertainties in the R magnitude
in addition to  the aperture correction  arise because of the
proximity,  within $\sim 4''$, of the much brighter  galaxy
(galaxy 174 in the cluster; Figure 1) to the northeast.

The $R$ magnitude reported here is somewhat brighter than the original
limit $r > 25.7$ mag of Persson et al.(1993).  The measurement at $R$
and limit at $r$ are entirely consistent  with the transformation
between $r$ and $R$ magnitudes given by Kent (1985).  The color
derived from the Keck data is  $R-K=7\pm0.5$ mag, and is dominated by
the uncertainty in the $R$ magnitude.  

There are no galaxies apparent closer than 4$''$ to CL~0939+4713 B ,
although  CL~0939+4713 A, as reported by Persson et al.(1993), is $\sim
7''$ away and has an $r-K$ color of 6.2 mag.  The Keck observations
give an $R$ magnitude of 24.75 mag for  CL~0939+4713 A, in excellent
agreement with the $r$ magnitude of 24.7 mag from  Persson  et al.

CL~0939+4713 B is very clearly resolved in the $I$ image with a
magnitude in a 2$''$ diameter beam of  $I = 23.26 \pm$0.07 mag. The
aperture correction to a 3$''$ diameter beam is $\sim -0.1$ mag
yielding an $I$ magnitude in a 3 $''$ diameter beam of $I=23.16
\pm0.15$ mag. The measured $I$ magnitude leads to a flux density a
factor of $\sim$3 greater than the flux density determined from the $i$
magnitude of Persson et al. (1993). The extended long wavelength
spectral response  in the $I$ measurement can account for some, but not
all, of this discrepancy.

The LRIS $R$ and $I$-band photometry is included in  Figure 3 and is
discussed below.

\smallskip

\section{Discussion}
\label{sec: Discussion}

The nature of the high Galactic latitude extremely red objects is
still  highly uncertain. Spectral energy distributions of old stellar
populations, and young, reddened galaxies are consistent with the
broad band observations for many of the extremely red objects, and
more detailed observations are required to distinguish between these
alternatives.   In the case of HR 10 (Hu and Ridgway, 1994), the
detection of emission lines and a submillimeter continuum  (Graham and
Dey, 1996, Cimatti et al. 1998, Dey et al. 1999) argue convincingly
that this is a comparatively ``young" galaxy whose colors are
substantially affected by reddening.

The observed $R-K$ color of CL~0939+4713~B places it among the
extremely red objects found in field surveys at $K$ (Hu and Ridgway
1994, Cohen et al. 1998, Barger et al. 1999, Thompson et al. 1999). 
The  spectral energy
distribution of Figure 3 allows  us to place constraints
on the redshift, the reddening,  and the evolutionary state of this
galaxy.  We first provide a qualitative view, considering two extreme
states, a very young and a very old galaxy,  and then introduce galaxy
models to deduce quantitative results.  We consider the possibility of
dust  in the nearby galaxy 174 affecting CL~0939+4713~B to be most
unlikely. The projected separation of the two objects, $\sim$30 kpc at
z=0.41 is  very large, so significant dust would have to present at a
very large distance from the center of that galaxy. Secondly,  the low
redshift (z=0.41) of galaxy 174 means that any reddening of the
observed infrared light in CL~0939+4713~B due to the intervening
galaxy would be affecting wavelengths not substantially different from
the observed wavelengths.  This would require substantially more dust
to produce an observable effect  than would be the case if the dust is
associated with the (presumably) much higher redshift source
CL~0939+4713~B.

If the galaxy is very old, the integrated light is dominated by G and
K stars.  The apparent break near 1.2\,$\mu$m then must be the
4000\,\AA\ break.  It cannot be the 2800\,\AA\ break, as this would
be weaker than the 4000\,\AA\ break, which  would then lie at
$\sim$1.7\,$\mu$m, and there is no sign of this feature in the
spectral energy distribution (SED) of Figure 3.   As described below,
we consider the possibility of the break being due to the Lyman limit
being most unlikely.  The redshift is thereby  constrained to lie near
z\,$=$\,2. The flat continuum observed  in the near infrared is then
the Paschen continuum of G/K stars, and since  the observed continuum
is not very red itself, the reddening must then be small.

If the galaxy is very young, there is no natural strong break in the
spectrum of hot stars except the Lyman limit. We rule out a Lyman limit
break because of the detection of the galaxy at $R$  and $I$, as well
as  considerations of plausible luminosities.  The break and apparent
slope must then be attributed to the effects of reddening.  The spectra
of local starbursts are smooth (Schmitt et al. 1997) and the reddening
is approximately  inversely proportional to wavelength so that there is
no way to reproduce the strong break seen at 1.2\,$\mu$m in the
spectrum.   In this case, there is no qualitative  match to the
observed SED with a normal reddening curve for any choice of redshift.

To make this comparison more quantitative, we fit the observed infrared
spectrum with the population synthesis models of Bruzual and Charlot
(1993, 1996).  Because the data are not of high signal to noise ratio,
there is limited information that can be derived from model fitting.
We fit instantaneous burst models of metal abundances $Z=Z_{\odot}$,
$Z=0.2Z_{\odot}$ and  a range of ages.  For each age model, the
redshift and  reddening (using the reddening law from Gordon and
Clayton, 1998) were varied to minimize the $\chi^2$ of the fit. The
models were fit only to the near-infrared data because the $R$ and $I$
data were qualitatively and quantitatively different from the grism
data. The consistency of the $R$ and $I$ data with the various model
fits was, however, checked. The results of the fits are given in Table
2.

The redshift of CL~0939+4713~B, determined in this manner, was
comparatively insensitive to the metal abundance,  age and reddening
of the model   and was established primarily by fitting the  drop in
observed flux below 1.2\,$\mu$m to the break in the continuum for the
model. Formally  the best fit redshift is consistently around
1.58, although the minimum reduced $\chi^2$ was 2.5 or greater. For 
a given model, the $\chi^2$ increases substantially at
z$<$1.55 and z$>$1.59, so  we take 1.55$<$z$<$1.59, or
z$=1.58+0.01/-0.03 $ as the result of our fitting.  Persson et
al. (1993) determined a redshift of 1.8 for this object based on
fitting their $r, i, J, H$ and $K$ data to Bruzual (1985) models.
Presumably the difference between the photometric redshift and that
determined here is fitting the 4000\AA\ break to the steep slope
observed at $\lambda <$ 1.2$\mu$m in the spectroscopic data presented
here.

While none of the models provides a statistically acceptable fit, the
$\chi^2$ parameter provides a quantitative measure consistent with the
visual  comparison of the relative goodness of fit of the different
models. The older models with solar metal abundances, $Z=Z_{\odot}$,
provide better fits to the data than do the models with abundances
$Z=0.2Z_{\odot}$. The break at 1.2\,$\mu$m is quite strong, and best
fit by old stellar populations with solar abundances.  Effects that
increase the restframe flux shortward of rest wavelength
$\sim$4000\,\AA\,  such as decreased UV opacity due to lower
metallicity or the presence of hot stars,  give lower quality fits to
the data.  The results of the fits to the $Z=Z_{\odot}$ models are
shown Figure 3.   In Figure 3 we have plotted the best fit models in
the observed frame  with $Z=Z_{\odot}$ for ages of $10^7, 10^8, 
2.9 \times 10^8, 10^9$
and $10^{10}$ yrs.   For the  models with ages of $10^9$ and $10^{10}$
yrs, the overall  continuum shape, both the slope at $\lambda > 1.2
\mu$m and the steep  drop to shorter wavelengths, are well fit with a
stellar continuum and  modest reddening.  In the two younger stellar
population models, the slope  of the continuum is set predominantly by
the reddening, with the redshift  of the Balmer discontinuity then
adjusted accordingly.  The younger age  population models produce too
much flux in the UV to match both the  continuum at $\lambda > 1.2
\mu$m and the drop to shorter wavelengths with the smooth reddening
curve adopted. The adoption of a different reddening law, such as that
of Calzetti (1997), does not modify these conclusions, only the
quantative amount of reddening, since over the wavelength range
appropriate to the fit the reddening is a power law A$_v\sim
\lambda^{-1}$ with no unique spectral features.

The models with a single initial burst of star formation lead to the
reddest colors for a given age, and thereby lead to the least
reddening required to fit a given age stellar population to the
observed spectrum.  We  use the model fits to constrain the formation
redshift of CL~0939+4713~B.  Because a given age single burst model is
redder than any model with ongoing star formation, this model fit to
the data will be the youngest population model consistent with the
data.  Thus, the single burst models are the most ``conservative" fits
to the observed spectrum of Figure 3, i.e. the youngest and least
reddened, and hence require  the lowest formation redshift to fit the
data.  More complex models,  e.g. exponentially decaying burst models,
would require star formation to have begun at a larger redshift.

We have included the $R$ and $I$ photometric fluxes in Figure 3.  Because 
of the wide bandpass of the two LRIS filters, and the
wide range in SEDs between the calibrating stars and the models, care
was taken to  properly account for the  shift in effective wavelength
in the different models, ie.  color correcting the predicted fluxes to
the scale set by the standards measurements. Because the range of
colors of standards is much less than the colors of the ERO, this by
necessity was done by calculation rather than observations. 

The models with  $t \ge 10^9$yr provide better fits to the infrared
spectrum than do the  younger models. At wavelengths $\le$ 1.2\,$\mu$m
the older models follow the observed drop in flux, while the younger
models  do not.  The observed  $R$-band flux is consistent with the
flux calculated  for the  $10^9$ yr age models. The models with ages of
$10^8$ and  $10^{10}$ years are also acceptable fits to the observed
$R$-band flux.  Clearly any single burst model of age substantially $<
10^8$ yrs produces significantly more flux than observed.   Including
the $I$
data suggest that the age is less than that suggested by the infrared
data. If the  model is constrained to pass through the measured $R$ and 
$I$ fluxes, a model with an age of 2.9 $\times 10^8$ years agrees with the
infrared data, although the quality of the fit to the infrared data alone is 
significantly worse that for the older models. 

If $z=$\,1.58 for CL~0939+4713 B the rest-frame absolute $B$ magnitude
is  $M_B= -$22.3 mag, uncorrected for extinction.  Mobasher, Sharples
and Ellis (1993) find $M_B^* \simeq -21.3$ mag for nearby galaxies, so
that the luminosity of CL~0939+4713 B is $\simeq$ 2.5$L^*$,  a
reasonable value for a massive galaxy at this redshift. A significant
uncertainty in the luminosity calculation is whether CL~0939+4713 B
suffers significant magnification due to its presence behind a
foreground rich cluster (Dressler and Gunn, 1992). While analysis of
the magnification due to the foreground cluster is beyond the scope of
this paper, Smail, et al. (1998) suggest a median (flux) magnification
factor in the line-of-sight to rich clusters including 0939+4713 (but
somewhat closer to the center of the cluster than represented by
CL~0939+4713~B) of a factor of 2.5.  Such a magnification factor would
serve to reduce the luminosity of CL~0939+4713~B to a present day $L^*$
galaxy if there is no significant extinction affecting the observed
flux.

As noted above, there is no evidence for  emission lines in the
spectrum of  CL~0939+4713~B.  The 3~$\sigma$ limit on the equivalent
width of any emission line in the spectrum of Figure 2 is
0.004\,$\mu$m. If those portions of the spectrum most affected by the
atmosphere and filter cut-offs are removed from the calculation of the
standard deviation of the flat continuum, the 3~$\sigma$ equivalent
width is decreased to 0.003\,$\mu$m.  This  corresponds to a limit on
the equivalent width of any emission line in the rest frame spectrum
from $ 0.4 - 1.0$\,$\mu$m of 12\,\AA\ if the redshift is given by our
fit to the continuum. In particular, at a redshift of 1.55-1.60,  the
3 $\sigma$ limit on the rest equivalent width of  H$\alpha$+[NII] is
$< 12$\,\AA\ .   This limit is more than a factor of 10 lower than the
strength of the H$\alpha$+[NII] line found in the near infrared
spectrum of HR 10 by Dey et al. (1999).  We can immediately say that
the emission line strengths of HR 10 and CL~0939+4713 B are very
different if our determination of the redshift of CL~0939+4713 B is
correct.

If we assume that the equivalent width of H$\alpha$ is 0.75 of the
equivalent width of H$\alpha$+[NII](Kennicutt, 1983), this corresponds
to a limit of  $< 9$\,\AA\ on the rest equivalent width of H$\alpha$.
We convert the equivalent width to a line luminosity using the
observed magnitude and a redshift z $\simeq$ 1.6 (and assuming no
magnification); using the relationship between H$\alpha$ luminosity
and star formation rate from Kennicutt, this corresponds to a limit of
$<9  M_{\odot}yr^{-1}$ on the star forming rate of  CL~0939+4713 B. In
terms of equivalent width, the limit of  $< 12$\,\AA\ on the
equivalent width of H$\alpha$+[NII] would place CL~0939+4713 B among
the less active disk galaxies in the local neighborhood (Kennicutt) or
the least active of infrared selected galaxies (Veilleux et al. 1995).
The limit on the H$\alpha$ equivalent width is consistent with
continuous star formation lasting for  more than $3 \times 10^8$ yrs
based on the models of Leitherer and Heckman (1995).  Thus there is
reasonably compelling evidence that CL~0939+4713 B is not a reddened,
young star forming galaxy, as is the case for HR 10.

Cohen, et al. (1999) have suggested that the objects found in the
Caltech Faint Galaxy Redshift Survey with $R-K> 5$ mag are old systems
that are not reddened, consistent with the interpretation of Spinrad et
al. (1997).  The identification of  CL~0939+4713~B as a distant galaxy
whose SED is not significantly affected by dust obscuration places it
in the category of potentially old galaxies at high redshift.

HR 10 and  CL~0939+4713 B have quite similar visual - infrared
colors. The detailed properties, i.e. morphology and infrared spectra
of these two systems are very different, leading to very different
pictures of the underlying galaxies. This suggests that  the objects
selected on the basis of red colors between visual and near infrared
wavelengths (EROs) represent a heterogeneous group, and a much larger
sample of these systems must be carefully studied to establish their
context in the high redshift universe.

\acknowledgments 

We thank W. Harrision for assistance with the infrared observations, Rob
Ivison, Len Cowie  and Amy Barger for obtaining the $I$ image, 
Marcin Sawicki for helpful
discussions and Dave Thompson for a careful reading of the manuscript
and several important suggestions.     The W.M. Keck Observatory is
operated as a scientific partnership between the  California Institute
of Technology, the University of California and the  National
Aeronautics and Space Administration. It was made possible by the
generous financial support of the W.M. Keck Foundation.  Infrared
astronomy at Caltech is supported by grants from the NSF and NASA.
This research has made use of the NASA/IPAC Extragalactic Database
which is operated by the Jet Propulsion Laboratory, Caltech, under
contract with NASA.

\clearpage


\clearpage
\begin{table}

\caption{ Photometry of CL 0939+4713 B}

\smallskip
\begin{tabular}{c c c  }   
\tableline\tableline
 Filter & mag (3$''$ diam. beam) & Ref   \\

\tableline

R  &   25.13$+$0.5/$-$0.3  & (1)  \\

r &    $>$25.7 3$\sigma$    & (2)    \\

i &   25.19$\pm$0.4  & (2)    \\
I &   23.16$\pm$0.15 & (1)   \\

J &   20.48$\pm$0.4  & (2)   \\
J &   20.36$+$0.37/$-$0.28 & (1)\\
H &   19.50$\pm$0.3 & (2) \\
H &   18.98$\pm$0.20 & (1) \\
K &   18.26$\pm$0.10 & (2) \\
K &   18.16$\pm$0.06 & (1) \\

\tablenotetext{ } {(1) This work}

\tablenotetext{ } {(2) Persson et al. (1993)}

\end{tabular}

\end{table}
\clearpage
\begin{table}


\caption{ Redshift and Reddening Resulting from the
Best Fits of Stellar Population Models to the Infrared 
Spectrum of CL 0939+4713 B}

\smallskip
\begin{tabular}{c c c c c }   
\tableline\tableline
 & & $Z_{\odot}$ &  & 0.2$Z_{\odot}$  \\
Age & z & $A_v$ & z & $A_v$ \\
 yr &   &  mag           &   &  mag    \\
\tableline

$1\times10^{10}$ &    1.578   & 0.1   & 1.578 & 0.5   \\

$1\times10^{9}$ &    1.582   & 1.1   & 1.582 & 1.4   \\

$2.9\times10^{8}$ &    1.588   & 1.9   & 1.586 & 1.9   \\

$1\times10^{8}$ &    1.589   & 2.3   & 1.582 & 2.2   \\

$1\times10^{7}$ &    1.589   & 1.8   & 1.586 & 2.4   \\

\tableline

\end{tabular}

\end{table}


\clearpage

\begin{figure}
\caption{  Images of CL 0939+4713 B  at $R$ and $K$, obtained using LRIS
on the Keck II  and NIRC on the Keck I telescopes  respectively.  The
field is 36 $''$ square for both filters, and the scales are
identical. North is up and east to the left.  CL 0939+4713 B and A are
both clearly seen in the $K-$band image, and are visible on the $R$-
band image. The location of CL 0939+4713 B is 9$^h$ 42$^m$56.8$^s$
+46$^o$57$'$48$''$ (J2000). 
 }
\label{fig: images of CL 0939+4713 B}
\end{figure}

\begin{figure}
\caption{ Spectrum of CL 0939+4713 B  plotted as flux density
$f_{\lambda}$ (Wm$^{-2}\mu$m$^{-1}$) vs. observed wavelength (in
$\mu$m). The hatched zones correspond the wavelength intervals that
have large corrections in the division by the G star caused by low
atmospheric transmission.
  }
\label{fig:observed spectra}
\end{figure}

\begin{figure}
\caption{  Spectrum of CL 0939+4713 B, plotted as log [flux/octave
$\lambda f_{\lambda}$ (W m$^{-2}$)] vs. observed wavelength (in
$\mu$m), and compared to models of  ages $10^7, 10^8, 2.9 \times 10^8,
10^9$  and $10^{10}$yr.  The infrared  data (filled circles) are
plotted at the resolution of the spectrum  and the infrared fluxes have
been adjusted to be consistent with the near infrared  photometry.  The
models are for solar metal abundance, and the redshift and reddening
corresponding to the values in Table 1. The models proceed from
youngest to oldest from top to bottom.  The $R$ and $I$ photometric
fluxes are included as filled  diamonds.  Because of the widths of the
band-passes, these fluxes depend on the assumed object spectrum. For
the models shown, the flux per octave ranges from -17.48 $< log[\lambda
f_{\lambda} (W m^{-2})]< $-17.36 ($I$) and -18.27 $< log[\lambda
f_{\lambda} (W m^{-2})]< $-17.94 ($R$).  The values calculated for a
Bruzual and Charlot (1993, 1996) model with an age of 2.9$\times 10^8$
yrs are plotted; the precepts outlined in the Explanatory Supplement to
the IRAS Catalogs and Atlases (Joint IRAS Science Team 1989) for color
corrections were followed. The effective wavelengths of the band were
chosen as 0.61 and 0.81~$\mu$m. The horizontal bars on the  $R$ and $I$
points indicate the full widths at half peak of the  filter and CCD
transmission.
  }
\label{fig:observed spectra}
\end{figure}

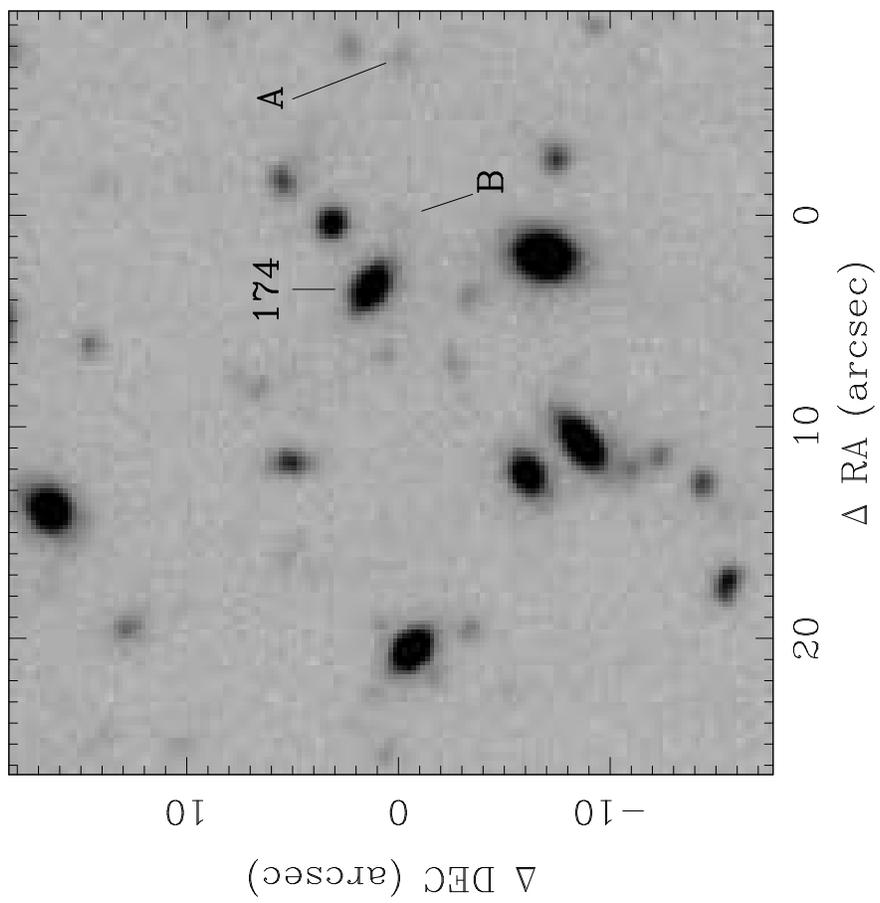
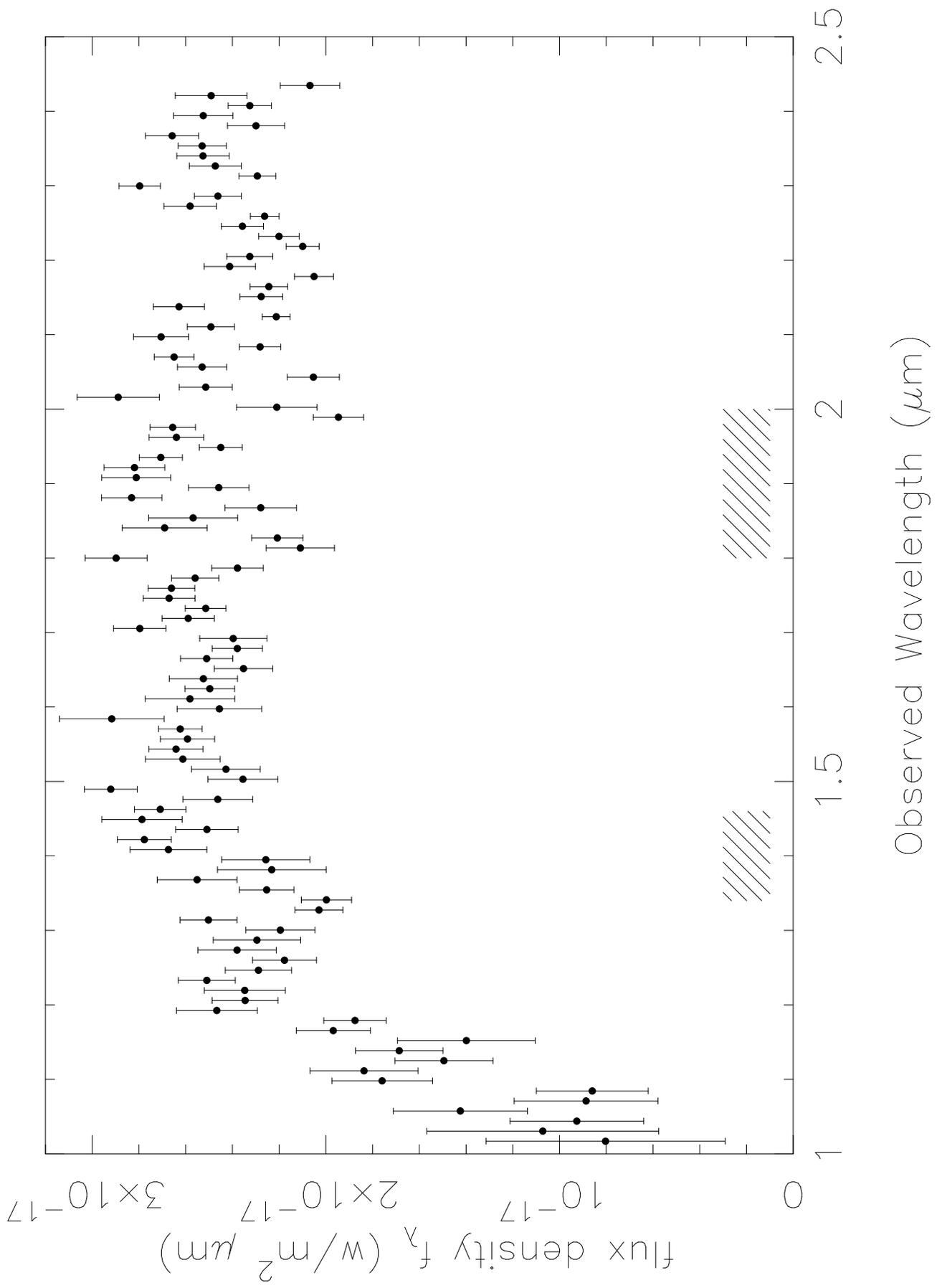
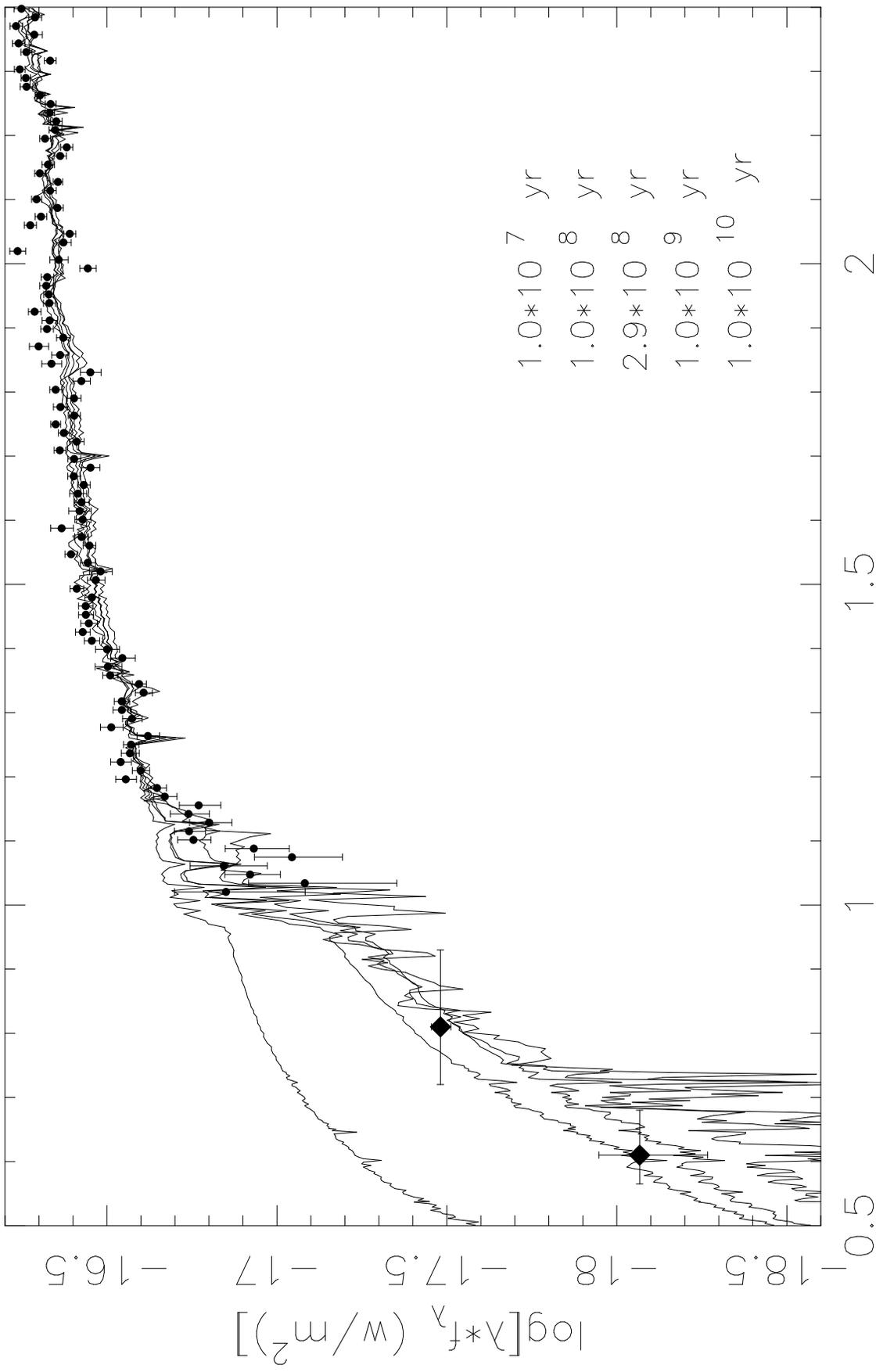
\end{document}